\documentclass[12pt]{article}
\usepackage{scicite}
\usepackage{times}

\usepackage{graphicx}
\usepackage{amsmath}

\newenvironment{sciabstract}{\begin{quote} \bf}{\end{quote}}

\newcounter{lastnote}

\title{Supersonic Gas Streams Enhance the Formation of Massive Black Holes in the Early Universe}

\author
{Shingo Hirano,$^{1,2\ast}$ Takashi Hosokawa,$^{2,3,4}$ Naoki Yoshida,$^{2,4,5}$ Rolf Kuiper$^{6}$\\
\\
\normalsize{$^{1}$Department of Astronomy, University of Texas, Austin, TX 78712, USA}\\
\normalsize{$^{2}$Department of Physics, School of Science, University of Tokyo,}\\
\normalsize{Bunkyo, Tokyo 113-0033, Japan}\\
\normalsize{$^{3}$Department of Physics, Kyoto University, Kyoto 606-8502, Japan}\\
\normalsize{$^{4}$Research Center for the Early Universe, University of Tokyo, Tokyo 113-0033, Japan}\\
\normalsize{$^{5}$Kavli Institute for the Physics and Mathematics of the Universe}\\
\normalsize{(World Premier International Research Center Initiative),}\\
\normalsize{University of Tokyo Institutes for Advanced Study, University of Tokyo,}\\
\normalsize{Kashiwa, Chiba 277-8583, Japan}\\
\normalsize{$^{6}$University of T$\ddot{\rm u}$bingen, Institute of Astronomy and Astrophysics,}\\
\normalsize{Auf der Morgenstelle 10, D-72076 T$\ddot{\rm u}$bingen, Germany}\\
\\
\normalsize{$^\ast$To whom correspondence should be addressed; E-mail: shirano@astro.as.utexas.edu}
}

\date{}

\begin{document}

\baselineskip16pt

\maketitle

\begin{sciabstract}
The origin of super-massive black holes in the early universe remains poorly understood.
Gravitational collapse of a massive primordial gas cloud is a promising initial process,
but theoretical studies have difficulty growing the black hole fast enough.
We report numerical simulations of early black hole formation starting from realistic cosmological conditions.
Supersonic gas motions left over from the Big Bang prevent early gas cloud formation
until rapid gas condensation is triggered in a proto-galactic halo.
A protostar is formed in the dense, turbulent gas cloud, and it grows by sporadic mass accretion until it acquires 34,000 solar masses.
The massive star ends its life with a catastrophic collapse to leave
a black hole -- a  promising seed for the formation of a monstrous black hole.
\end{sciabstract}

\clearpage
Recent discoveries of super-massive black holes (SMBHs) at redshift $z \sim 7$,
when the universe was just 5\% of its present age,
pose a serious challenge to the theory of black hole (BH) formation and evolution\cite{mortlock11}.
The physical mechanisms that form the BH seeds and drive their growth are not yet known
but must explain how an initial seed BH can reach
a mass of 10 billion times that of the Sun ($M_\odot$) within 1 billion years after the Big Bang.
It is thought that the mass growth is a self-regulating process,
limited by the so-called Eddington rate that is proportional to the BH mass; therefore.,
starting from a massive BH holds a key to the rapid formation of SMBHs.

A previously-proposed physical model posits that
an early BH forms through direct gravitational collapse of a large primordial gas cloud
with a mass of about $10^5\,M_\odot$\cite{haiman13}.
However, the model must invoke several critical conditions, such as
the formation of a bright galaxy in the immediate vicinity\cite{dijkstra08,visbal14,latif15a,regan17}.
The overall occurrence rate of such a supposedly rare event in a cosmological volume remains rather uncertain.
We present an ab initio simulation of the formation of massive BHs and
examine the importance of the supersonic streaming motion
between dark matter and ordinary baryonic matter in the early universe\cite{tseliakhovich10}.
In regions with a large streaming velocity, gas condensation -- and hence star formation -- is suppressed
until a deep gravitational potential is generated by a clump of dark matter with mass $10^7\,M_\odot$\cite{fialkov12},
where similar conditions to that of the conventional direct-collapse model might be realized\cite{tanaka14}.

We performed cosmological hydrodynamics simulations of early structure formation that incorporate the relative bulk velocity.
We selected target dark-matter halos whose dynamical properties were consistent with those of the host halos of SMBHs\cite{wang13,SOM}.
We then re-generated the initial conditions for three zoom-in simulations with
a fixed streaming velocity of three times the root mean square (RMS) value\cite{SOM}.
In the following material, we concentrate on describing one case, Run-B,
which has $\sigma_8 = 1.2$, corresponding to $1.5$ times the cosmic mean value, where $\sigma_8$ is the RMS density fluctuation in a sphere of radius 8 Mpc.
We also performed two other runs:
Run-A, with a larger fluctuation amplitude $\sigma_8 = 2$; 
and Run-C, starting from a different realization of the density field with $\sigma_8 = 1.2$ (table~S1).

At early epochs in Run-B, the streaming motions prevented gas cloud collapse
in small-mass dark-matter halos with $10^5\,M_\odot$\cite{stacy11,greif11},
which would otherwise host the first generation of stars\cite{tegmark97}.
Dark-matter halos grew hierarchically through mergers and accretion, and
a massive halo with $2.2 \times 10^7\,M_\odot$ was assembled
when the cosmological redshift was $30.5$ (Fig.~1A).
The host halo was about 100 times as heavy as
than that found in previous simulations without the streaming motions (table~S1).
The strong gravity trapped the streaming gas (Fig.~1B), and
the excess momentum generated a wide wedge-shaped structure (Fig.~1C).

The trapped gas was firstly shock-heated to a temperature of $\sim 10$,$000$\,Kelvin (K) and
the temperature soon decreased to $\sim 8000$\,K by atomic-hydrogen (H) cooling (Fig.~2).
The compressed gas further cooled through molecular-hydrogen (H$_2$) emission and
formed a dense and cold gas cloud (fig.~S1).
The cloud was still supported against gravitational collapse
by thermal pressure and also by turbulence induced by streaming motions.
For such a dynamically supported gas with density $\rho$, the Jeans mass is given by
$M_{\rm Jeans} = (\pi / 6) (v_{\rm eff}^3 / G^{3/2} \rho^{1/2})$,
where $G$ is the gravitational constant and
$v_{\rm eff} = \sqrt{c_{\rm s}^2 + v_{\rm bc}^2}$,
with $c_{\rm s}$ and $v_{\rm bc}$ denoting
the sound speed and
relative velocity between baryon and cold dark-matter components, respectively\cite{stacy11,naoz13}.
The assembled cloud became gravitationally unstable when its mass exceeded $26$,$000\,M_\odot$
at particle number density $4000\,{\rm cm^{-3}}$ and gas temperature $400$\,K (Fig.~1D and fig.~S2).
The rapidly inflowing gas was accumulated on the contracting cloud and produced a massive and dense envelope (fig.~S3A).
During the subsequent collapse, a fully molecular cloud formed by efficient H$_2$ formation
via three-body reactions at $10^8\,{\rm cm^{-3}}$\cite{palla83},
causing the molecular core to collapse further.
After this epoch, the temperature evolution of the core was similar to that in the ordinary primordial star formation\cite{yoshida08},
but the whole cloud contracted more rapidly, and
the instantaneous gas mass accretion rate exceeded $1\,M_\odot\,{\rm year}^{-1}$ in the outer part of the cloud (figs.~S3 and S4).

The conventional model of direct BH formation\cite{bromm03,tanaka14} assumes that
the gas evolves nearly isothermally at a high temperature of $\sim 8000$\,K.
That model invokes a few necessary conditions so that
the gas can evolve on a high-temperature track to maintain a large gas mass accretion rate in the later accretion phase.
Often, photo-dissociation of hydrogen molecules by a nearby galaxy is assumed\cite{visbal14}.
The massive gas cloud in our simulation cooled first by H cooling  and then by H$_2$ cooling.
The temperature drop at a density of $100\,{\rm cm^{-3}}$ and
the onset of H$_2$ cooling have been considered as a failure of direct collapse,
possibly leading to cloud fragmentation\cite{latif15b},
but our simulations show that the cloud collapse continues in a highly dynamical manner.

We stopped our cosmological simulation
when a tiny protostar with a mass of $\sim 0.01\,M_\odot$ was formed at the gas cloud center.
We then turned to a three-dimensional gravito-radiation-hydrodynamic simulation
coupled with a self-consistent modeling of the stellar evolution\cite{SOM}
in order to follow directly the complex interplay between the gas mass accretion and
the stellar radiation feedback that reduces the mass accretion rate (Fig.~1, E to H).
When the mass accretion rate is greater than a critical value of $0.04\,M_\odot\,{\rm year}^{-1}$,
the excess entropy carried by rapidly accreting gas causes the stellar envelope to inflate.
A star with an extended envelope has a low effective temperature of $\sim 6000$\,K and hence
does not emit copious amounts of ultra-violet (UV) photons.
The resulting radiative feedback on the surrounding gas is weak and cannot halt the gas accretion.
The central star in our simulation grew rapidly to a mass of $50\,M_\odot$ in the first 2000 years (Fig.~3A).

Gas accretion onto the star occurred sporadically owing to the combined effect of
the temperature structure of the infalling gas and mass accretion through the circumstellar disk,
where the gravitational torque dominated the angular momentum transport\cite{inayoshi14,latif15b}.
The mass accretion rate fluctuated around the critical value of $0.04\,M_\odot$\,year$^{-1}$ for the first $300$,$000$\,years (Fig.~3B).
During the early phase, the star contracted and its UV emissivity increased temporarily
when the accretion rate fell below the critical value (Fig.~3, C and D).
However, gravitational fragmentation of the cloud and subsequent merging of the fragments caused accretion bursts,
and the star quickly recovered to have an extended envelope\cite{sakurai15,sakurai16,hosokawa16}.
Because successive accretion bursts occurred at short time intervals in the late accretion phase,
the stellar envelope remained extended
because there was insufficient time for it to contract.

The hydrogen molecules formed along the dense filament (Fig.~1E) were dissociated
by the far-UV radiation emitted by the central star.
A bi-polar ionized atomic hydrogen (H~\textsc{ii}) region appeared temporally (red contour in Fig.~1F)
due to the increasing intensity of the stellar UV radiation.
The surrounding dense gas filament fragmented to yield a gas clump (Fig.~1G),
but it did not cool and contract further because it contained few hydrogen molecules, and
gradually approached the central star, avoiding the bipolar H~\textsc{ii} region.
The clump finally reached the center (Fig.~1H),
activated an accretion burst (Fig.~3B) onto the star, and weakened the radiative feedback.
After the protostellar mass reached $\sim 10$,$000\,M_\odot$,
it evolved as a stable super-giant protostar without radial contraction and continued growing steadily up to $34$,$000\,M_\odot$.
Run-A and Run-C also showed the formation of massive primordial stars with $100$,$000$ and $4400\,M_\odot$ at the simulation end.
Such very massive accreting stars undergo direct-collapse to produce equally-massive BHs
owing to the general relativistic instability or exhaustion of nuclear fuel, depending on the accretion rate\cite{SOM,umeda16}.

A $34$,$000\,M_\odot$ BH formed at $z = 30.5$ must grow at about $55$\% of the canonical Eddington rate until $z = 7.1$ to acquire a mass of $2 \times 10^9\,M_\odot$,
matching the estimated mass of the SMBH observed in a luminous quasar\cite{mortlock11,marziani12,wu15}.
The number density of the intermediate mass BHs formed by this mechanism is 
estimated to be
$\sim 1$ per cubic comoving gigaparsec\cite{SOM},
which is similar to the abundance of the observed SMBHs\cite{willott10}.

Unlike previous studies of direct-collapse BH formation, our simulations did not assume additional conditions,
such as the existence of strong radiation sources\cite{omukai01} and/or
high-speed collisions of gas clouds\cite{inayoshi15} (see also supplementary text).
Other than the primeval density fluctuations,
whose statistical properties are well understood from both theory and observation,
the only newly introduced element in our study is the relative velocity between the dark-matter and baryonic components.
The baryonic streaming motions are intrinsically generated in the early universe
according to the standard model of structure formation\cite{tseliakhovich10}.
Therefore, our ab initio cosmological simulations show a viable formation path for massive BHs.

\clearpage
\paragraph*{Acknowledgments:}
The numerical calculations were carried out on Cray XC30 at Center for Computational Astrophysics, National Astronomical Observatory of Japan, and COMA at Center for Computational Sciences, University of Tsukuba.
This work was supported by Japan Society for the Promotion of Science KAKENHI grants 14J02779 to S.H.,; 25800102, 15H00776, and 16H05996 to T.H.,; and by Japan Science and Technology Agency CREST JPMHCR1414, MEXT Priority Issue 9 on Post-K Computer to N.Y. 
R.K. acknowledges financial support via the Emmy Noether Research Group on Accretion Flows and Feedback in Realistic Models of Massive Star Formation funded by the German Research Foundation (DFG) under grant no. KU 2849/3-1. 
R.K. is also affiliated with the Max Planck Institute for Astronomy in Heidelberg, Germany. 
The source code for our customized versions of GADGET and PLUTO, along with links to other open- source packages and input files necessary to reproduce our simulations, are available at https://shirano.as.utexas.edu/SV.html.
Snapshots of the output data from each simulation are available on the same page.

\paragraph*{Supplementary Materials} \

\noindent www.sciencemag.org/content/357/[issue]/[page]/supppl/DC1\\
\noindent Materials and Methods\\
\noindent Supplementary Text\\
\noindent Figs. S1 to S9\\
\noindent Table S1\\
\noindent References ({\it 30}--{\it 53})\\

\noindent 31 August 2016; resubmitted 16 March 2017; accepted 23 August 2017

\clearpage
\begin{figure}[!ht]
\includegraphics[width=\textwidth]{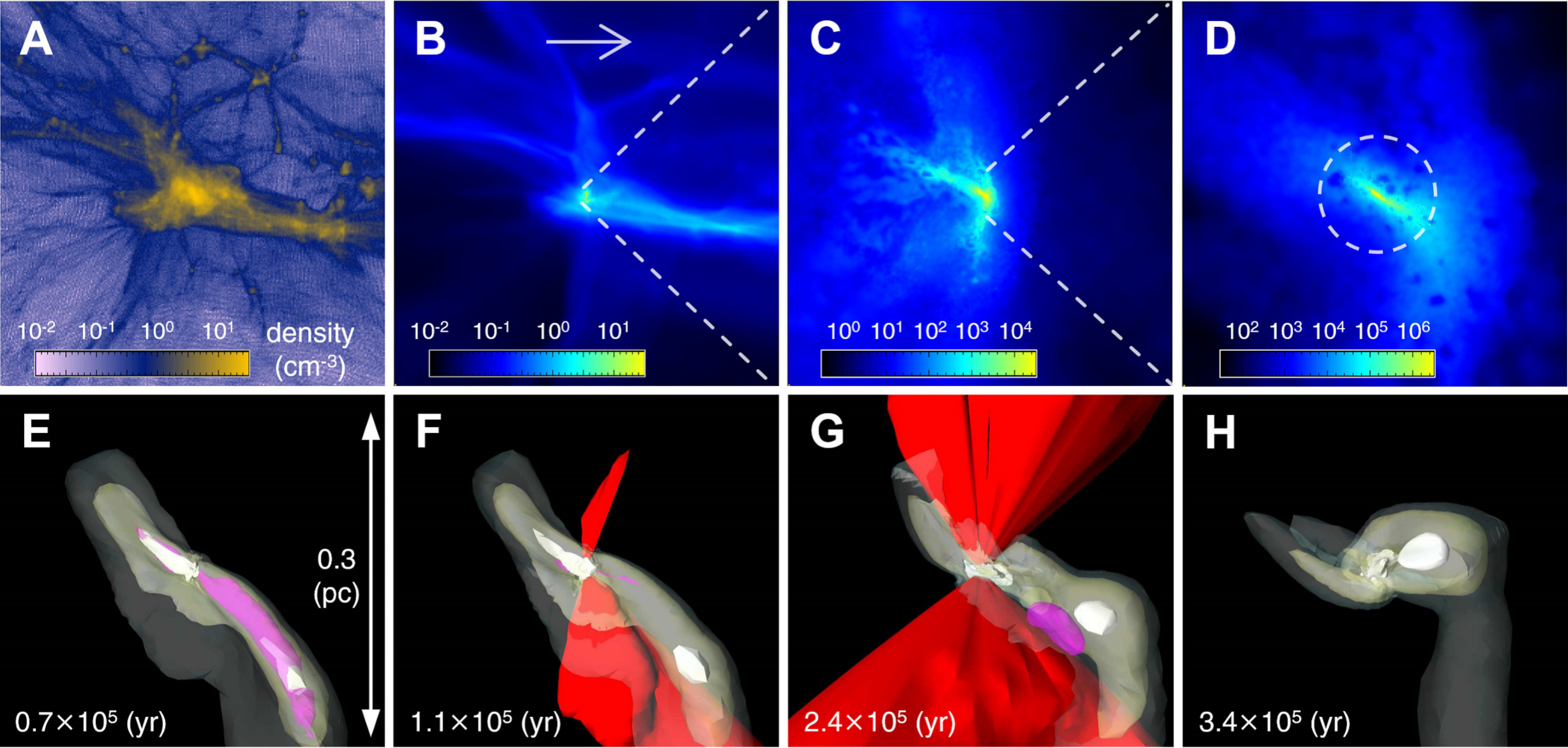}
{
{\bf Fig.~1. Large-scale density distribution and the structure around an accreting protostar.}
({\bf A}) Projected density distribution of dark-matter component around the star-forming cloud at redshift $z = 30.5$ in Run-B.
The box size is $2500$\,parsec (pc) on a side.
The virial mass of the main dark matter halo located at the center is $2.2 \times 10^7\,M_\odot$.
({\bf B} to {\bf D}) Projected density distribution of the gas component
in regions of $2500$, $100$, and $10$\,pc on a side, from left to right.
The horizontal arrow in ({\bf B}) shows the direction of the initial supersonic 
gas stream.
The dashed circle in ({\bf D}) indicates the Jeans length,
within which the cloud is gravitationally unstable given its mass of $26$,$000\,M_\odot$.
({\bf E} to {\bf H}) Evolution of the temperature and density structure in the protostellar accretion phase
after the protostar formation.
Colored in white, red, and magenta are
the iso-contours of gas density (at $10^6$, $10^5$, and $3 \times 10^4\,{\rm cm}^{-3}$),
photoionized hydrogen abundance with $\ge 50$\% (H~\textsc{ii} region), and
the number fraction of hydrogen molecules with $\ge 0.2$\%.
}
\end{figure}

\clearpage
\begin{figure}[!ht]
\includegraphics[width=\textwidth]{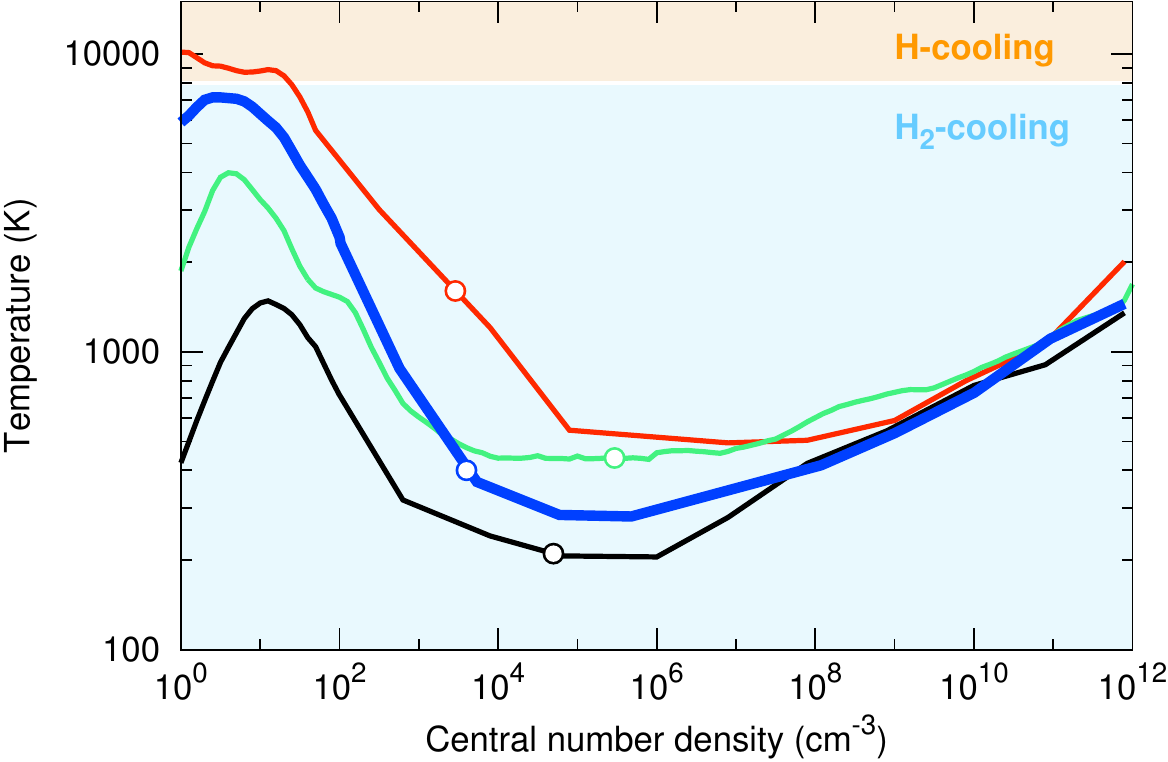}
{
{\bf Fig.~2. Thermal evolution of the collapsing cloud.}
The red, blue, green, and black lines show our Run-A, B, C, and Ref, respectively.
In each run, the cloud became Jeans-unstable at the points marked by the circles.
The background colored regions are distinguished by the major coolant: 
atomic hydrogen (H) and molecular hydrogen (H$_2$).
In all the cases, H$_2$ cooling operated at densities greater than $\sim 100\,{\rm cm^{-3}}$.
}
\end{figure}

\clearpage
\begin{figure}[!ht]
\begin{center}
\includegraphics[width=7.5cm]{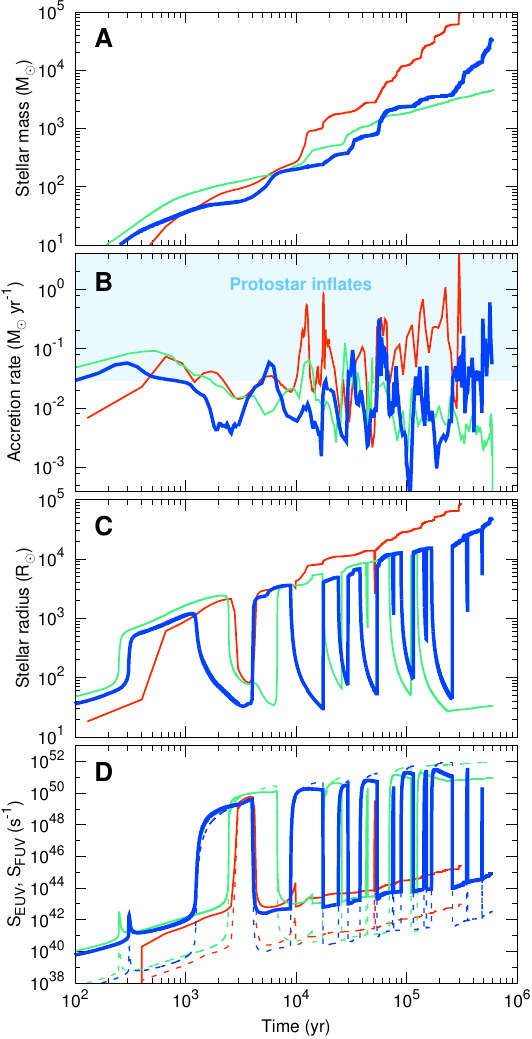}
\end{center}
{
{\bf Fig.~3. Time evolution of stellar properties.}
({\bf A}) Stellar masses,
({\bf B}) mass accretion rates,
({\bf C}) stellar radii, and
({\bf D}) extreme ultra-violet (EUV) (with $h\nu \ge 13.6$\,eV; solid) and
far ultra-violet (FUV) (with $11.2\,{\rm eV} \le h\nu \le 13.6$\,eV; dashed) emissivities,
where $h$ is the Planck constant and $\nu$ is the frequency of a photon.
The red, blue, and green lines represent our Run-A, B, and C, respectively.
When the mass accretion rate was larger than the critical value of $0.04\,M_\odot$\,year$^{-1}$ (Run-A),
the star expanded continuously during the accretion phase, and the FUV/EUV radiation could not halt gas accretion.
When the gas accretion rate fluctuated around the critical value 
due to sporadic accretion (Run-B),
the stellar radius and the resulting UV emissivity, repeated rises and falls\cite{sakurai15,sakurai16,hosokawa16}.
The accretion rate stayed well below the critical value for a sufficiently long time in its final phase in Run-C and
the protostar entered the zero-age main-sequence.
}
\end{figure}

\clearpage

\begin{center}

\baselineskip24pt

\ \\

{\Large Supplementary Materials for}\\ \ \\

{\large Supersonic Gas Streams Enhance the Formation of \\Massive Black Holes in the Early Universe\\ \ }

\baselineskip16pt

Shingo Hirano, Takashi Hosokawa, Naoki Yoshida, Rolf Kuiper\\ \ \\

correspondence to: shirano@astro.as.utexas.edu\\ \ \\
\end{center}

\baselineskip14pt
\paragraph*{This PDF file includes:} \ \\

\hspace{0.5cm} Materials and Methods

\hspace{0.5cm} Supplementary Text

\hspace{0.5cm} Figs. S1 to S9

\hspace{0.5cm} Table S1

\hspace{0.5cm} References

\clearpage
\subsection*{Materials and Methods}

\noindent \underbar{Cosmological Simulations}

We perform a set of three-dimensional cosmological simulations to study
the primordial star formation by incorporating the effect of baryonic streaming motions.
The streaming motions can be included in a straightforward manner
because the distribution of the streaming velocity is coherent
over a length scale of a few comoving megaparsecs (cMpc), which is larger than regions
that contain atomic-hydrogen cooling halos of interest to super-massive black hole (SMBH) formation\cite{tseliakhovich10}.
Because the cosmological streaming velocity (SV) is not correlated with the local overdensity\cite{ahn16},
the initial streaming direction can be set arbitrarily.
We introduce the initial relative velocity between baryonic and cold dark matter components $v_{\rm bc}$
as a constant uniform velocity along one axis into the cosmological initial conditions generated at redshift $z_{\rm ini} = 499$.
The root-mean-square value of $v_{\rm bc}$ is $\sigma_{\rm bc,rec} = 30\,{\rm km\,s^{-1}}$
at the epoch of cosmological recombination $z_{\rm rec} = 1089$\cite{PLANCK13XVI}.
We set $3\sigma_{\rm bc}(z_{\rm ini})$ as the initial value in our cosmological simulations,
i.e. $v_{\rm bc}(z_{\rm ini}) = 90\,{\rm km\,s^{-1}}\,(1+z_{\rm ini})/(1+z_{\rm rec}) \sim 41\,{\rm km\,s^{-1}}$.

We run four simulations starting from different initial conditions:
three cases (Run-A, B, and C) with the same SV but with different density fluctuations, and
one reference case (Run-Ref) without SV.
The initial amplitudes of the density fluctuation are
$\sigma_8 = 2.0$ for Run-A and
$\sigma_8 = 1.2$ for Run-B and Run-C
which are higher than the observational constraint $\sigma_8 \sim 0.8$\cite{PLANCK13XVI}.
For the main case Run-B, we selected a target dark matter halo
whose central velocity dispersion is $\sim 160$\,km\,s$^{-1}$ at $z=7$,
being consistent with the estimated value of the host galaxies of observed SMBH\cite{wang13}.
Run-A is expected to represent a rare, high-density peak.
The reference run is initiated from the fiducial cosmological initial conditions with $\sigma_8 = 0.8$.
We adopt the standard $\Lambda$-Cold Dark Matter cosmology with the Planck cosmological parameters\cite{PLANCK13XVI}:
matter density $\Omega_{\rm m} = 0.3086$,
baryon density $\Omega_{\rm b} = 0.04825$,
dark energy density $\Omega_{\Lambda} = 0.6914$ in units of the critical density,
the Hubble constant of $h = 0.6777$, and
the primordial spectral index $n_{\rm s} = 0.96$.

We generate the initial conditions in a (10\,$h^{-1}$\,cMpc)$^3$ cosmological volume using MUSIC\cite{hahn11}.
The cosmological simulations are performed by using
the parallel $N$-body / Smoothed Particle Hydrodynamics (SPH) code GADGET\cite{springel05b}
suitably modified for primordial star formation\cite{hirano15a}.
We use a hierarchical zoom-in technique to generate
the initial conditions for the target halos with higher mass and spatial resolutions.
The particle masses of the dark matter and baryonic components in the zoom-in regions are $16.4$ and $3.0\,M_\odot$, respectively.
During the cloud collapse, we pose a strict refinement criterion that
the local Jeans length is always resolved by $15$ times the local SPH smoothing length.
We achieve this by progressively increasing the spatial resolution using the particle-splitting technique\cite{kitsionas02}.
In all runs, we stop the cosmological SPH simulation
when the hydrogen number density $n_{\rm H}$ at the cloud center reaches $10^{12}\,{\rm cm}^{-3}$.
At this point, we define that a protostar is formed at the maximum density site.
Although a protostar is actually to be formed when the central density exceeds $10^{20}\,{\rm cm}^{-3}$\cite{yoshida08},
the time difference between the two epochs is very small.
In all runs, we find a single protostar at the end point of our SPH simulations.

To characterize the main star-forming gas clouds, we define two relevant length
and mass scales, the virial and Jeans scales.
The virial length scale is defined as the radial distance from the center
within which the mean density (including dark matter)
is $200$ times the cosmic mean value.
The Jeans scale is defined as the radius where
the ratio of the enclosed mass to the local effective Jeans mass,
$M_{\rm Jeans} = (\pi / 6) (v_{\rm eff}^3 / G^{3/2} \rho^{1/2})$,
takes its maximum (Fig.~S2).
To evaluate the effective Jeans mass, we consider the additional dynamical
support generated by the streaming velocity,
$v_{\rm eff} = \sqrt{c_{\rm s}^2 + v_{\rm bc}(z)^2}$
where $v_{\rm bc}(z) = 3\sigma_{\rm bc,rec} (1+z) / (1+z_{\rm rec})$.
The actual values for the main gas clouds are summarized in Table~S1.

From the mass of the host dark matter halo evaluated at the time
when a dense gas cloud is formed within it, we can estimate the
number density of the early black holes (BHs) as follows.
Since the cosmological streaming velocity and the local over-density are not correlated at the length scales of
our interest here\cite{ahn16}, 
the number density of the intermediate-mass BHs formed as in the present study can be estimated by multiplying the number density of the host dark matter halos and the probability distribution of the streaming velocity. 
Using the halo mass function of\cite{reed07} with our standard cosmological parameters\cite{PLANCK13XVI}, 
we obtain the abundance of dark matter halos with mass $> 10^7\,M_\odot$ to be $\sim 5000$ per cubic cGpc (comoving gigaparsecs) at $z = 30$, and $\sim 8.9 \times 10^5$ per cubic cGpc at $z = 25$.
The probability of such halos being located in regions with more than 
$3\sigma$ streaming motions is $\sim 5.9 \times 10^{-6}$ and that for more than 
$2.7 \sigma$ is $\sim 6.9 \times 10^{-5}$\cite{tseliakhovich11,schauer17}.
Multiplying these numbers yields the BH number density of $\sim 0.03$--$61$\,cGpc$^{-3}$.\\

\noindent \underbar{Gravito-Radiation-Hydrodynamic Simulations}

The accretion process of the new-born protostar is followed by using
the three-dimensional hydrodynamic code PLUTO\cite{mignone07}
augmented by self-gravity\cite{kuiper11} and radiation transfer\cite{kuiper10}.
This gravito-radiation-hydrodynamics framework was further coupled to the protostellar evolution code STELLAR\cite{yorke08}
suitably modified to study primordial star formation\cite{hosokawa16}.
We use the gas opacity for a gas with primordial composition.

The basic numerical settings are adopted from the previous study\cite{hosokawa16} with the following modifications.
First, we use a combination of direct integration and an analytic prescription to follow the evolution of the central star.
We numerically follow the stellar evolution using the STELLAR code
until the stellar mass reaches $\sim 100\,M_\odot$.
Afterward, we switch to an analytic prescription. Our analytic model is based on and calibrated by the numerical results\cite{hosokawa16}.
Such an analytic prescription gives accurate results after the stellar mass exceeds $\sim 100\,M_\odot$,
when a protostar is either in the super-giant phase with a high accretion rate of $>0.04\,M_\odot\,{\rm yr}^{-1}$,
or in the Kelvin-Helmholtz contracting phase toward the main-sequence with a lower accretion rate\cite{sakurai15,sakurai16}.
With this treatment, we can avoid tuning the numerical convergence parameters many times
to construct very accurate stellar structure under variable mass accretion.
The computational cost is reduced and then we can perform the radiation-hydrodynamic simulation for a long time
until we can determine the final mass of the star.

Second, we eliminate a constraint that existed in our previous studies.
Earlier, we assumed that the star is spatially fixed at the center of the collapsing cloud, and thus
the gravitational attraction of the star by the non-spherically symmetric distribution of the surrounding gas was nor properly incorporated.
In our simulations presented here, the density distribution is rather filamentary, particularly in Run-B (Fig.~1C).
Hence, it is necessary to eliminate the unphysical constraint.
To this end, we solve the equations in the frame co-moving with the star.
The star remains in the center of the spherical coordinate system
(which allows a fast and accurate computation of radiative transfer as well as gravitational forces),
but the motion of the gas in the computational domain is adjusted by a pseudo-force in the co-moving frame.
Explicitly, we calculate the gravity force vector exerted on the coordinate origin from the surrounding gas,
by summing up the contributions from all the computational cells.
We then add to the gas an inertia force, which is just the opposite of the calculated gravitational force.

Finally, we gradually extend the central sink radius, $R_{\rm sink}$,
throughout the evolution from the initial value of $R_{\rm sink} = 116$\,au (astronomical unit).
We do this by simply removing the inner three cells roughly every $10^5$\,years.
The number of radial grids $N_{\rm r}$ thus decreases.
In Run-A and Run-B, for instance, we start the simulation with $N_{\rm r} = 106$ ($R_{\rm sink} = 116$\,au) and
end with $N_{\rm r} = 91$ ($R_{\rm sink} = 624$\,au),
in Run-C we end with $N_{\rm r} = 103$ ($R_{\rm sink} = 162$\,au).
This allows us to follow very long-term evolution over several times $10^5$\,years
after the birth of a protostar, with available computational resources.

Aside from those modifications, we otherwise adopt the calculation settings used in the previous work\cite{hosokawa16}.
The initial conditions of the gravito-radiation-hydrodynamic simulations are taken from snapshots of the cosmological simulation.
For that, the data from the three-dimensional (3D) cosmological SPH simulation are remapped onto 3D spherical meshes as follows.
The simulation volume is $\{ R_{\rm min}, R_{\rm max} \}$ = $\{ 115.6~{\rm au}, 1.73 \times 10^7\,{\rm au} = 83\,{\rm pc} \}$.
The number of meshes is initially $\{ N_{\rm r}, N_{\rm phi}, N_{\rm psi} \}$ = $\{ 106, 32, 64 \}$.
The physical quantities are averaged for each grid cell with weighting over SPH smoothing length.
The coordinate origin is set to be at the location of the local maximum density point and
the mass flow over the associated inner sink cell of the spherical coordinate system represents the stellar accretion rate.
We solve for the evolution of a single star within the sink.
We set the polar axis of the grid system to
the direction of the mean angular momentum vector of the gas enclosed within $1$\,pc from the origin.
With this choice, the circumstellar disk remains perpendicular to the polar axis of the spherical grids in Run-A and Run-C,
where the cloud is relatively spherical without strong deformation by the streaming velocity.
However in Run-B, where the cloud is stretched to have the filamentary shape,
the angular momentum vector of the accreting gas, or the rotational axis of the emerging circumstellar disk,
frequently changes during the accretion phase.

Regarding the cell number per Jeans length, we have always spatially resolved it with at least $12$ cells\cite{hosokawa16}.
Our simulations are able to follow the large-scale gravitational fragmentation within a star-forming gas cloud.
In fact, we see the formation and migration of such a fragment in Run-B;
the circumstellar filamentary cloud yields a fragment, which falls toward the central star and is soon accreted.
The accretion burst and the subsequent expansion of the stellar surface prevents
the further expansion of the H~\textsc{ii} region in the late evolutionary stage (Figs.~1E to 1H and 3).
Therefore, the complex interplay between cloud fragmentation and
the protostellar feedback is followed consistently in our 3D calculations.

With the limited spatial resolution, however, our simulations do not perfectly follow
the small-scale fragmentation within the circumstellar disk.
If the finer structure in the inner disk (within the sink radius) were resolved,
fragmentation could happen more frequently\cite{stacy16}.
The disk fragmentation is normally followed by inward migration and merger with the central star,
through which the central star grows\cite{greif12,hosokawa16}.
Interestingly, the highly variable accretion and associated short accretion bursts make it easier for a protostar to inflate;
the UV radiation feedback against the accretion flow tends to be weaker\cite{sakurai15,sakurai16,hosokawa16}.

We stop our gravito-radiation-hydrodynamics simulations when the final fate of the protostar is determined.
Run-A is stopped when the growing protostellar mass reaches $100$,$000\,M_\odot$.
The central core quickly becomes unstable via the general relativistic instability and collapses to produce a massive BH with $\sim 100$,$000\,M_\odot$\cite{umeda16}.
We stop Run-B when the protostellar mass is $34$,$000\,M_\odot$.
At this time, we can assume that the stellar mass will not increase substantially
because the accretion rate is already below $0.1\,M_\odot\,{\rm yr}^{-1}$.
Since the star has already evolved for over $6\times 10^5$\,years, it does not
grow in mass by a large factor even if the gas accretion continues during the remaining lifetime. At the end of its evolution,
the super-massive star with a mass greater than $3 \times 10^4\,M_\odot$ collapses and leaves a massive BH remnant.
We stop Run-C when the mass growth is halted by the stellar radiation feedback.
The stellar mass is determined to be $4$,$400\,M_\odot$. 
It is smaller than in Run-A and Run-B, 
but is still sufficiently large for the star to produce and 
leave an intermediate mass BH with mass $\sim 4$,$400\,M_\odot$.\\

\subsection*{Supplementary Text}

\noindent \underbar{Gas cloud collapse and mass accretion rate}

Our simulations show rapid gravitational collapse of a massive gas cloud under a large streaming velocity (SV).
The resulting gas mass accretion rate onto the central protostar is also very large.
Although these features appear similar to those expected in the conventional direct-collapse model, there is one unique effect caused by SV.
We identify a physical reason for the large accretion rate in Run-B.
The SV generates gas random motions, turbulence, during the assembly of 
the host dark halo and also during the contraction of the gas cloud.
The random motions act as an effective non-thermal pressure against
gravitational collapse, and thus
increase the effective Jeans mass.
Furthermore, when the assembled gas cloud finally undergoes run-away collapse, 
the instantaneous inflow rate of the surrounding gas is very large, 
exceeding $1\,M_\odot\,{\rm yr}^{-1}$. 
This inflow rate is even larger that expected for the direct-collapse model 
under strong ultra-violet (UV) radiation.

To compare directly the physical properties of collapsing gas clouds,
we have re-run cosmological simulations with the same initial conditions 
as for Run-B, but without streaming motions and with a strong UV radiation.
We assume a blackbody spectrum of Population II stellar sources with the effective temperature of $T_{\rm eff} = 10^4$\,K.
We have performed two simulations with different UV radiation intensity $J_{21}$ that is the intensity at the Lyman-Werner bands
normalized in units of $10^{-21}$\,erg\,s$^{-1}$\,cm$^{-2}$\,Hz$^{-1}$\,sr$^{-1}$:
Run-J300 with $J_{21} = 300$ and
Run-J1 with $J_{21} = 1$.
Run-J300 can be considered to be a typical direct-collapse case, where
a massive cloud gravitationally contracts with having a nearly constant gas temperature $8$,$000$\,K,
whereas Run-J1 corresponds to a failed direct-collapse, where
a massive gas cloud cools via H$_2$ emission.
We choose the latter value of $J_{21}$ because the resulting thermal evolution is close to that of our Run-B with SV.
Hence, differences between Run-B and Run-J1, if any, are attributed to some other effect(s) than the thermal evolution.

Naively, we expect that a gas cloud collapses in essentially the same manner in 
Run-B and in Run-J1,
because the gas thermal evolution is indeed similar (Fig.~S5B).
However, with SV, the gas cloud cannot collapse gravitationally
until the cloud's self-gravity overcomes the effective pressure support provided by
the gas thermal pressure and non-thermal pressure owing to turbulent motions that are generated by the excess SV.
In other words, in the SV run, the gas cloud remains dynamically hot when it is
thermally so cold that its thermal pressure support against gravity is weak.

Figure~S6 shows that the gas cloud in Run-J1 contracts nearly spherically.
In contrast, in Run-B, the gas cloud is strongly deformed by the streaming motion;
it is squashed in the horizontal direction in the figure.
The excess SV provides an effective dynamical support against self-gravity
of the massive gas cloud (Fig.~S4). There is another cosmological effect associated 
with this. With the non-thermal support in Run-B,
the gas cloud can collapse only at a later epoch than in Run-J1,
when the host halo has grown to be $14$ times more massive (Table~S1).
This is because, when the halo virial temperature reaches $8$,$000$\,K
at $z = 43.8$, the gas in Run-J1 cools and contracts quickly through H-cooling,
but the gas in Run-B is still supported effectively by SV.
The gas cloud in Run-B finally collapses at $z = 30.5$,
when the host halo mass has grown to be $2.2 \times 10^7\,M_{\odot}$.
Within the massive host halo, the gas cloud is already 
highly concentrated before it becomes gravitationally unstable (Fig.~S5).

Interestingly, the apparently rapid collapse of the gas cloud in Run-B is caused
by a combination of physical processes as follows.
First, the gas cloud gathered by a deep dark matter gravitational potential is supported dynamically by random motions that are generated by SV (Figs.~S4 and S7).
The gas cools via H$_2$-line emission and becomes gravitationally unstable
when the collected mass exceeds $26$,$000\,M_\odot$.
The rapidly inflowing gas is accumulated and the cloud becomes denser (Fig.~S3A).
When the gas density reaches $n_{\rm H} \sim 10^8$\,cm$^{-3}$,
the three-body reactions increase the H$_2$ fraction 
and hence the H$_2$ cooling rate.
This triggers the final collapse of the cloud core (Fig.~S3A).
The dense gas envelope around the core is eventually accreted,
to sustain the large accretion rate in later accretion phases.

We have performed a gravito-radiation-hydrodynamic simulation for Run-J1 
with the same settings as for Run-B.
The final stellar masses are $227\,M_\odot$ for Run-J1, and $34$,$000\,M_\odot$ for Run-B (Fig.~S8).
The mass difference reflects the different evolution and 
collapse dynamics under SV and UV,
despite the similarity in the gas thermal evolution.
The net effect of SV is seen most clearly in the instantaneous mass infall rate $\dot{M}$ (Fig.~S5D).
At the enclosed mass $M_{\rm enc} > 300\,M_{\odot}$, the gas infall rate is much larger in Run-B.
At $M_{\rm enc} \sim 10^4\,M_{\odot}$, there is an order of magnitude difference
in $\dot{M}$ despite the similar thermal evolution between Run-B and Run-J1 (Fig.~S5B).
The streaming velocity $v_{\rm bc}$ enhances the mass infall rate as
\begin{equation}
\dot{M} \sim M_{\rm J}/t_{\rm ff} \propto v_{\rm eff}^3 \sim (c_{\rm s}^2 + v_{\rm bc}^2)^{3/2}\, ,
\tag{S1}
\end{equation}
where $M_{\rm J}$ is the Jeans mass,
$t_{\rm ff}$ is the free-fall time, and
$c_{\rm s} \propto T^{1/2}$ is the sound speed.
When the gas cloud collapses at $z = 30.5$ in Run-B,
the streaming velocity is
$v_{\rm bc}(z) = 3\sigma_{\rm bc,rec}\,(1+z)/(1+z_{\rm rec}) = 90\,{\rm km\,s^{-1}}\,(1+30.5)/(1+1089) = 2.6$\,km\,s$^{-1}$.
This is slightly larger than the sound-speed in the gas, 
$c_{\rm s} \sim 2$\,km\,s$^{-1}$.
The effective velocity is larger by a factor of $1.52$ over $c_{\rm s}$,
and then the accretion rate is expected to be 
a factor of $1.52^3 = 3.51$ times larger. 
In addition to this, there is a large difference in the
mean gas density (Fig.~S5A) that accounts for the extremely large $\dot{M}$ shown in Fig.~S5B. 
The rapid decrease of the accretion rate in Run-J1 at $t \sim 20$,$000$\,years is not just a random event, 
but is caused by the strong radiation feedback from the central star (Fig.~S8). 
With the lower accretion rate in Run-J1, the central star contracted when its mass exceeded $200\,M_{\odot}$ and the surface temperature increased, to cause strong radiation feedback. 
This finally quenched the growth of the central star in Run-J1.

The degree of random turbulent motions is shown in Fig.~S4,
where we compare the gas infall velocity, the sound speed, and the local velocity dispersion. Clearly, strong turbulence has developed in the
outer part of the cloud\cite{greif11}.
At the enclosed mass greater than $1$,$000\,M_{\odot}$,
the gas infall velocity is larger than the local sound speed and
progressively approaches the local velocity dispersion at large masses.

The instantaneous mass infall rate at the Jeans scale, $\dot{M}_{\rm Jeans}$,
can be used as a proxy of the actual mean accretion rate and hence of the final stellar mass\cite{hirano14}.
Fig.~S9 shows $\dot{M}_{\rm Jeans}$ as a function of the gas temperature for all of our runs.
For comparison, we also plot $\dot{M}_{\rm Jeans}$ for
ordinary primordial star formation simulations as gray dots\cite{hirano14,hirano15a}.
Despite the large temperature difference, the infall rate 
in Run-A is almost the same as in
Run-J300 that is a direct-collapse case with strong UV radiation with $J_{21} = 300$.
Run-J1 is close to ordinary primordial star-forming clouds\cite{hirano14},
which can be understood from its thermal evolution, but Run-B and Run-C 
with similar gas temperatures have much larger $\dot{M}$.
Fig.~S9 clearly shows that SV boosts the gas infall rate.

The role of turbulence has been discussed in the study of present-day 
star-formation that predicts rapid gas infall in the formation sites of massive stars\cite{mclaughlin97,mckee02,mckee03}.
With strong turbulence, a gas cloud can be supported by non-thermal pressure 
even when its mass is much larger than the thermal Jeans mass.
Detailed radiation-hydrodynamic simulations of present-day massive 
star formation show that the collapse of a turbulent core is actually 
delayed \cite{krumholz07}.
The gas cloud in our runs with SV, in particular in Run-B,
evolves similarly to the turbulent core model.
The cloud is initially supported by non-thermal pressure,
and continues gathering a large amount of gas from the surrounding.
When its mass exceeds the effective Jeans mass, 
the cloud collapses in about a free-fall time (Equation S1).

Finally, the large host halo mass in Run-B can be understood as follows. 
To first order, the formation of dark matter halos is not affected 
by baryonic motions nor by UV radiation.
Hence the difference is actually in the epoch when the gas can cool and collapse.
The threshold halo mass for gas cloud collapse is given by
\begin{equation}
M_{\rm th} = \frac{v_{\rm circ}^2 R}{G}\, ,
\tag{S2}
\end{equation}
where $v_{\rm circ}$ is the circular velocity of the halo and 
$R$ is the halo radius.
Under baryonic streaming motions, the effective threshold mass is obtained
by replacing $v_{\rm circ}^2$ with
$v_{\rm circ,0}^2 + (\alpha v_{\rm bc})^2$
with $v_{\rm bc} = v_{\rm bc,rec} (1+z)/(1+z_{\rm rec})$.
We then obtain the $z$-dependence of
\begin{equation}
M_{\rm th} \propto \left[(v_{\rm circ,0})^2 \frac{1}{1+z} + \left( \frac{\alpha v_{\rm bc,rec}}{1+z_{\rm rec}} \right)^2 (1+z) \right]^{3/2}\, ,
\tag{S3}
\end{equation}
where $v_{\rm circ,0}=3.7\,{\rm km\,s^{-1}}$ and $\alpha = 4.0$ are
best fit parameters\cite{fialkov12} to the results of cosmological simulations\cite{stacy11,greif11}.
The streaming velocity decreases with redshift, but $M_{\rm th}$ remains roughly constant at $z = 30$ -- $40$.
Therefore, when the halo in Run-B grows to reach the threshold value (for cooling and collapse) at $z=30.5$,
its mass is much larger than the halo mass at $z=43.8$ in Run-J1.

For Run-J1 and Run-J300 presented here,
we do not model the formation of the radiation source(s).
We set the radiation intensity $J_{21}$ just as a parameter for 
the purpose of comparing a variety of simulations.
At redshift $z>30$, the radiation source likely consists of Population III stars with $T_{\rm eff} = 10^5$\,K 
for which the critical radiation intensity for direct collapse is 
about a hundred times higher than for Population II stars\cite{chon16}.
Adopting a more realistic spectral energy distribution, rather than the simple black-body spectrum, 
would yield an even higher critical radiation intensity\cite{sugimura14}. 
Therefore, it is unlikely that 
strong radiation sources exist already at $z = 43.8$ (Run-J1) nor at $z = 37.6$ (Run-J300) 
which can provide the necessary UV radiation intensity at the site of the primordial gas cloud.
Realistic cosmological simulations show that
the radiation-driven direct-collapse
occurs in very rare halos forming at low redshifts\cite{chon16}.
The resulting accretion rate is found to be as high as $\sim 0.1\,M_{\odot}\,{\rm yr}^{-1}$\cite{latif15c}.

\clearpage
\begin{figure}[!ht]
\begin{center}
\includegraphics[width=0.7\textwidth]{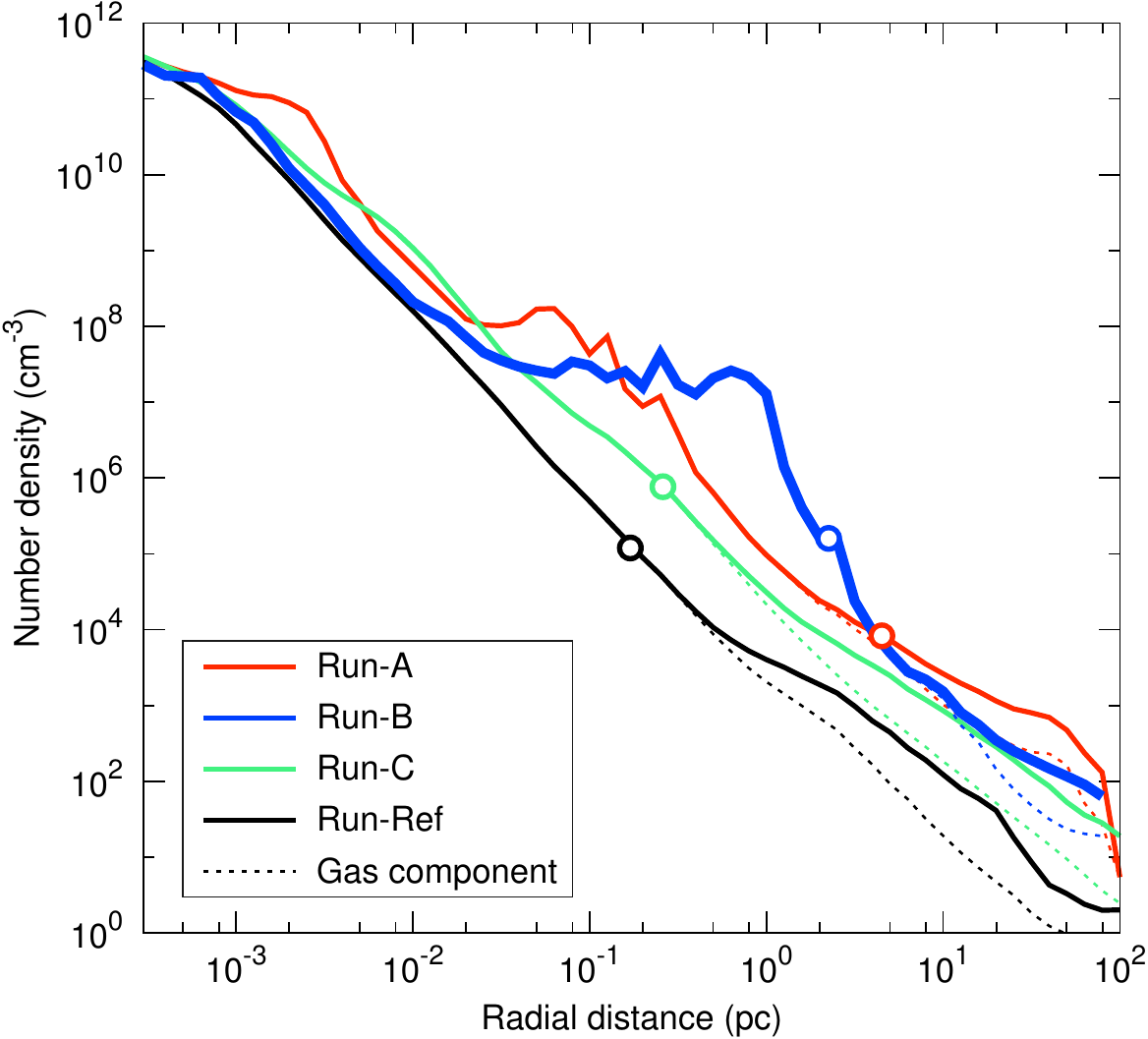}\\
\end{center}
{
{\bf Fig.~S1. Radial density profiles at the end of our cosmological simulations when the central gas density reaches ${\bf 10^{12}\,{\bf cm}^{-3}}$.}
The solid (dashed) lines are for the total (gas) density.
The circles indicate the radius within which the gas is gravitationally 
unstable (Fig.~S2) at the final output for each model.
The star-forming clouds in Run-A and Run-B are denser than in the reference run.
}
\end{figure}

\clearpage
\begin{figure}[!ht]
\begin{center}
\includegraphics[width=0.7\textwidth]{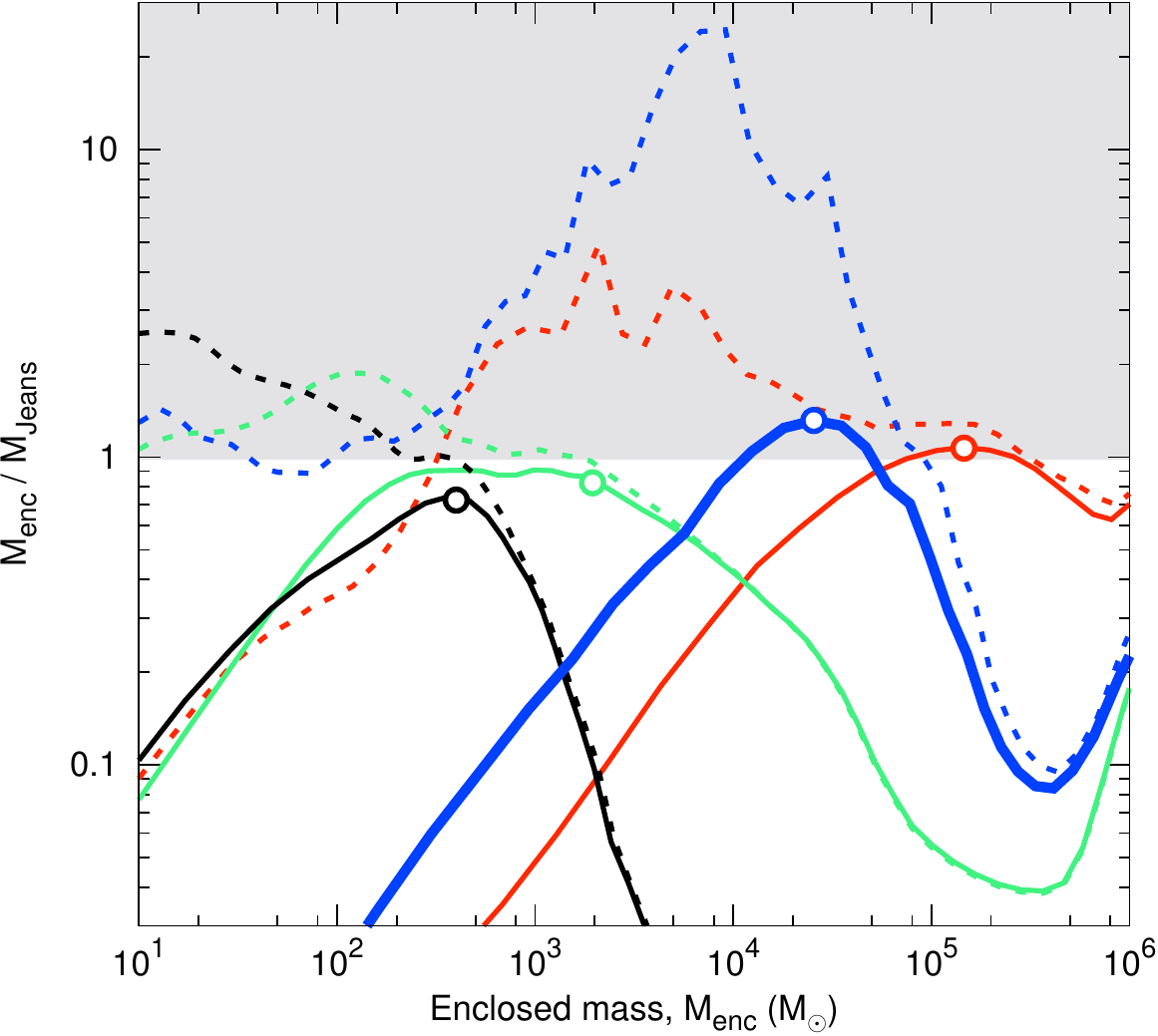}\\
\end{center}
{
{\bf Fig.~S2. Characteristic index of the gravitational instability of star-forming clouds.}
The horizontal axis, enclosed mass $M_{\rm enc}$, represents
a mass coordinate from the inner to the outer radius of the star-forming cloud.
The red, blue, green, and black lines show Run-A, B, C, and Ref, respectively.
The circles indicate the Jeans unstable mass for each cloud.
The solid lines show the same quantities at the first time when the 
clouds become gravitationally unstable,
$M_{\rm enc}/M_{\rm Jeans} \geq 1$ (with gray background),
whereas the dotted lines show the profiles at the end of the cosmological simulation.
}
\end{figure}

\clearpage
\begin{figure}[!ht]
\begin{center}
\includegraphics[width=1.0\textwidth]{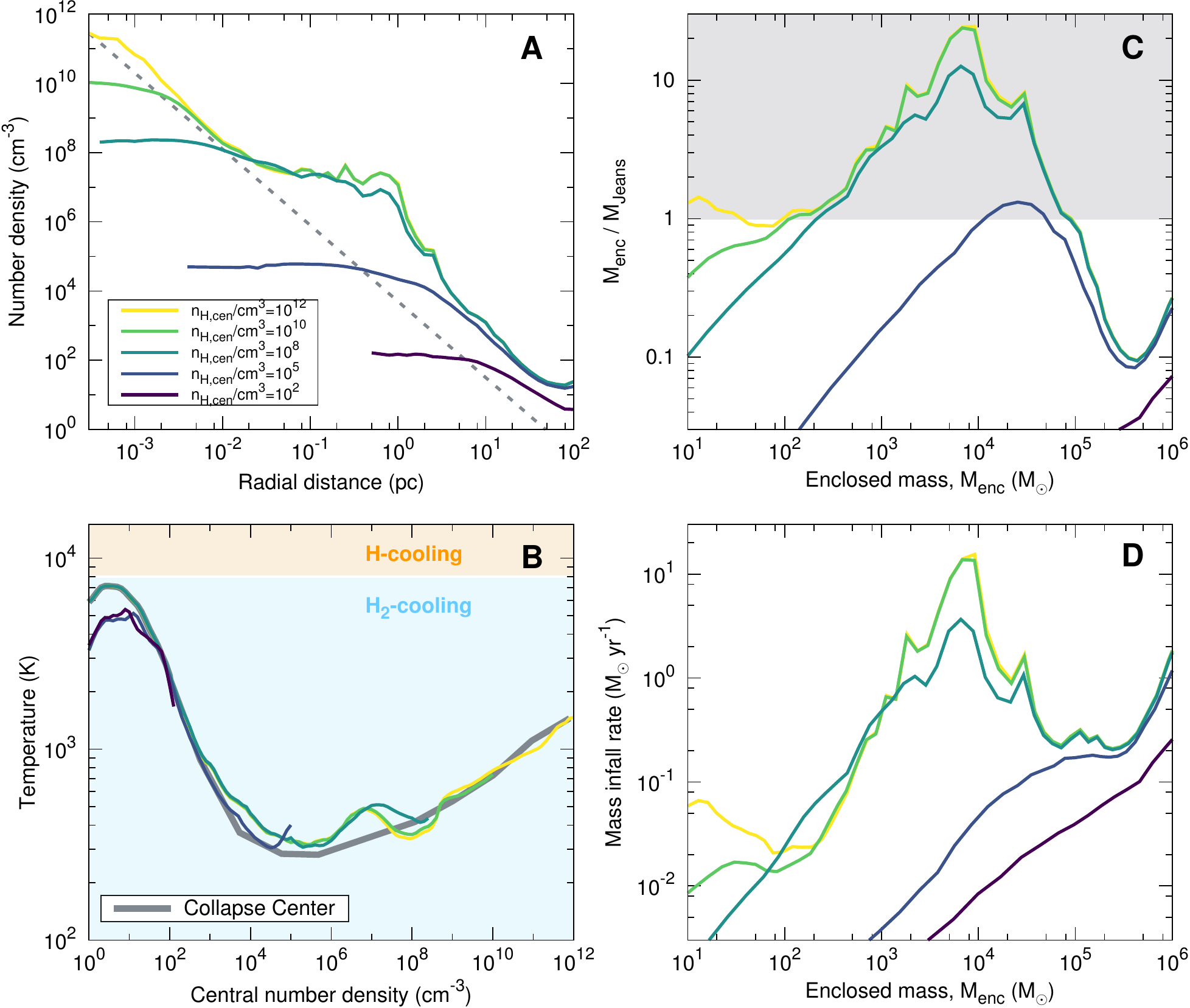}\\
\end{center}
{
{\bf Fig.~S3. Properties of the gravitationally collapsed gas cloud in Run-B.}
({\bf A}) Radial density profiles,
({\bf B}) thermal evolution,
({\bf C}) gravitational instability, and
({\bf D}) instantaneous mass infall rate
at the central gas density $10^{2}$, $10^{5}$, $10^{8}$, $10^{10}$, and $10^{12}\,{\rm cm}^{-3}$.
The dashed line in {\bf A} shows a characteristic power-law density distribution  
of primordial gas clouds,
$5$,$000\,{\rm cm}^{-3}\,(R/{\rm pc})^{-2.2}$\cite{omukai98}.
The gray line in {\bf B} shows the thermal evolutionary track of the collapsing center.
}
\end{figure}

\clearpage
\begin{figure}[!ht]
\begin{center}
\includegraphics[width=0.7\textwidth]{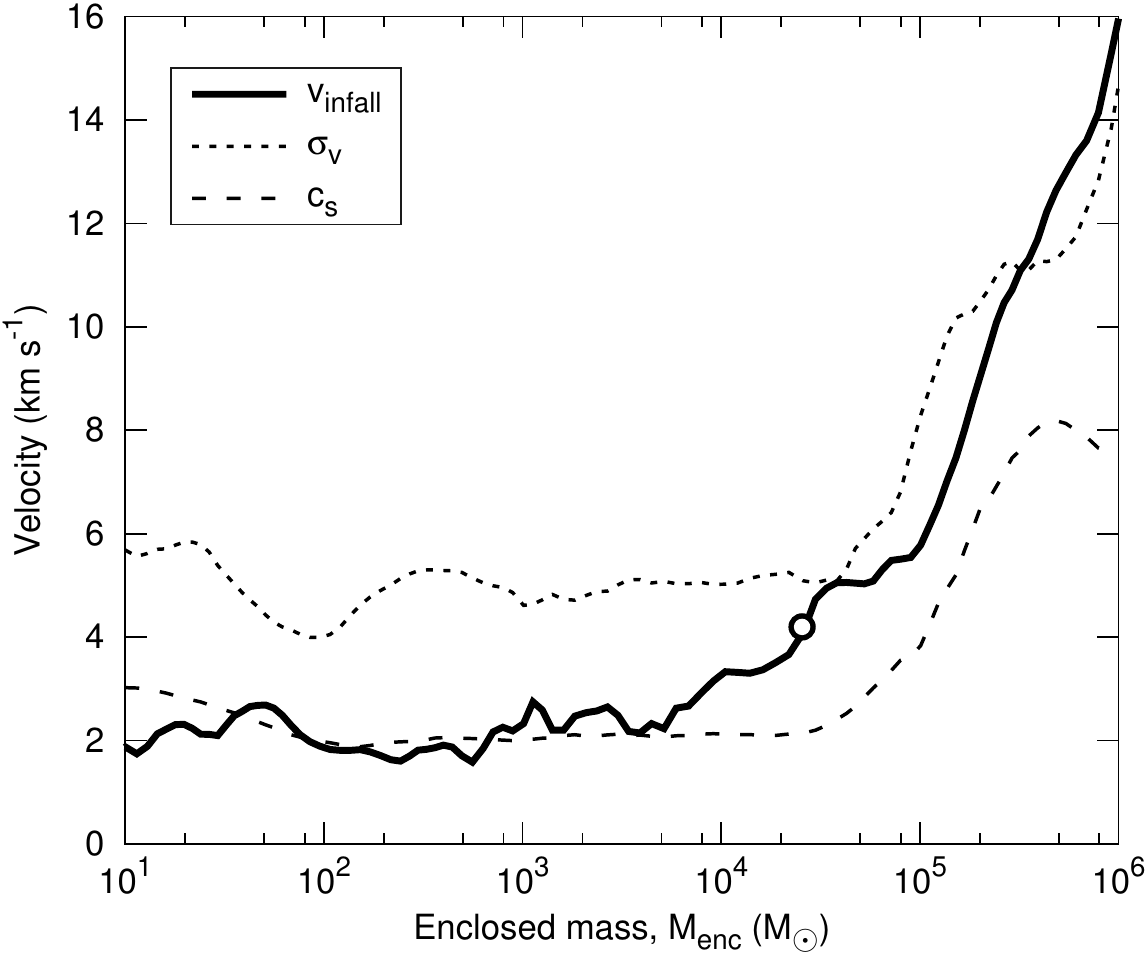}\\
\end{center}
{
{\bf Fig.~S4. Velocity profiles as a function of the enclosed mass when the central gas density reaches ${\bf 10^{12}\,{\bf cm}^{-3}}$ for Run-B.}
The solid, dotted, and dashed lines represent
the radial infall velocity, velocity dispersion, and sound speed, respectively.
The circle indicates the effective Jeans mass, where
the infall velocity is actually larger than the sound speed.
}
\end{figure}

\clearpage
\begin{figure}[!ht]
\begin{center}
\includegraphics[width=1.0\textwidth]{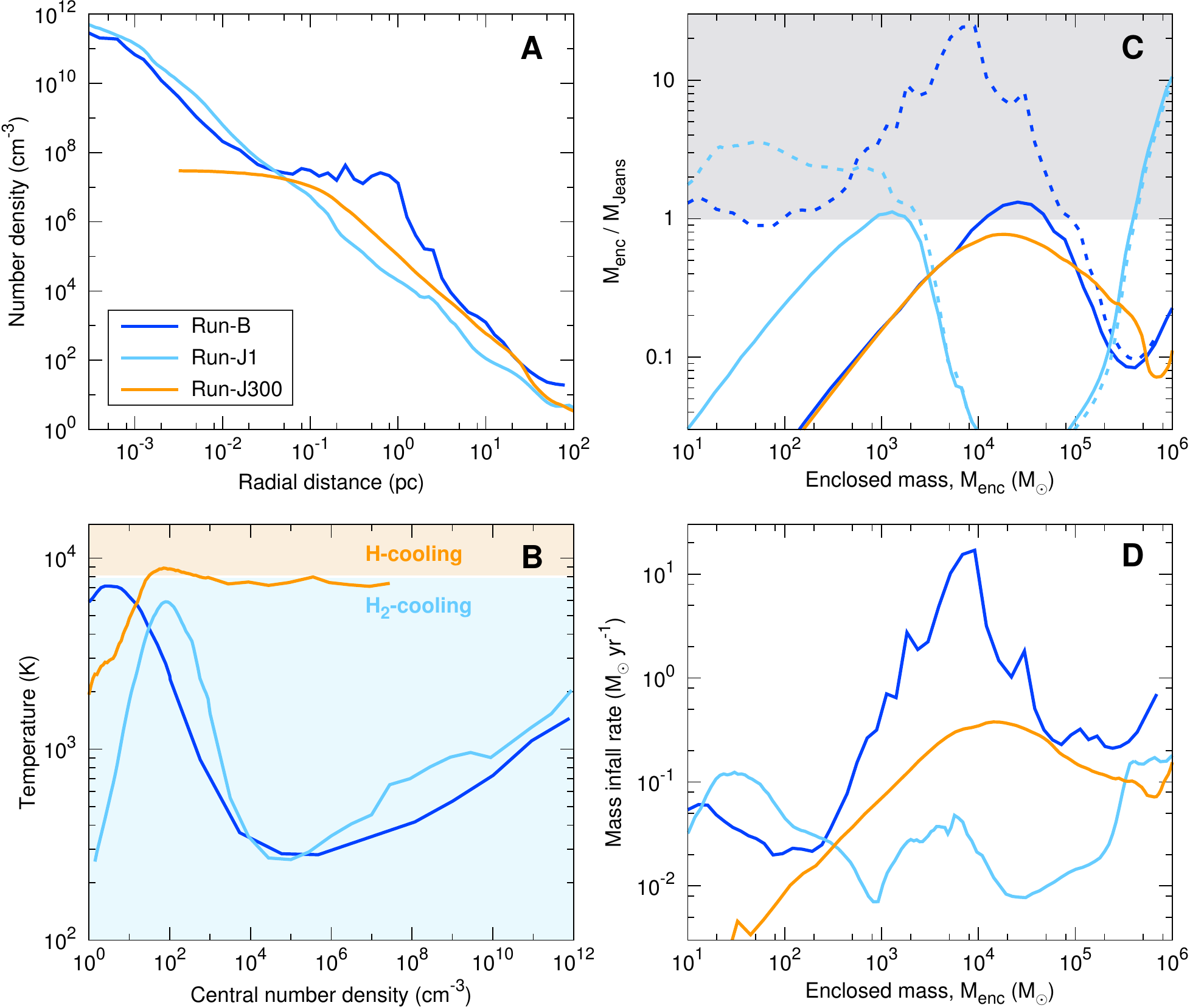}\\
\end{center}
{
{\bf Fig.~S5. Properties of gravitationally collapsed clouds.}
({\bf A}) Radial density profiles,
({\bf B}) thermal evolution,
({\bf C}) gravitational instability, and
({\bf D}) instantaneous mass infall rate when the central gas density reaches $10^{12}\,{\rm cm}^{-3}$.
The blue line are for Run-B.
The cyan and orange lines show
results for Run-J1 (with UV radiation background $J_{21} = 1$) and Run-J300 ($J_{21} = 300$), respectively.
}
\end{figure}

\clearpage
\begin{figure}[!ht]
\begin{center}
\includegraphics[width=1.0\textwidth]{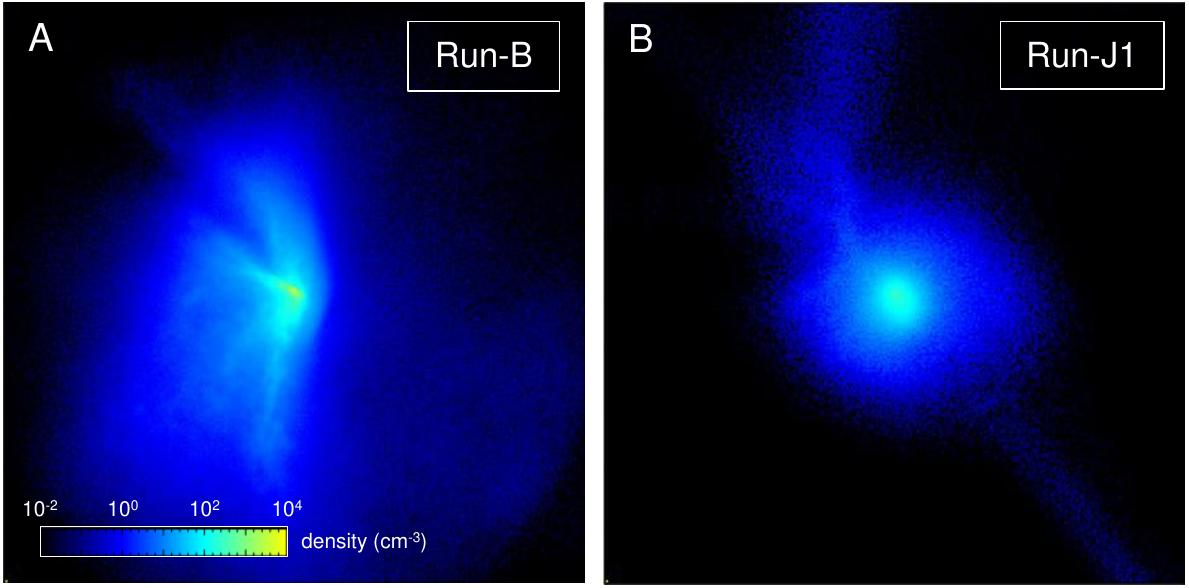}\\
\end{center}
{
{\bf Fig.~S6. Projected gas density distribution for Run-B (A) and Run-J1 (B) in a region of 300\,pc on a side.}
({\bf A}) The initial supersonic stream was set in the horizontal direction from left to right.
}
\end{figure}

\clearpage
\begin{figure}[!ht]
\begin{center}
\includegraphics[width=0.7\textwidth]{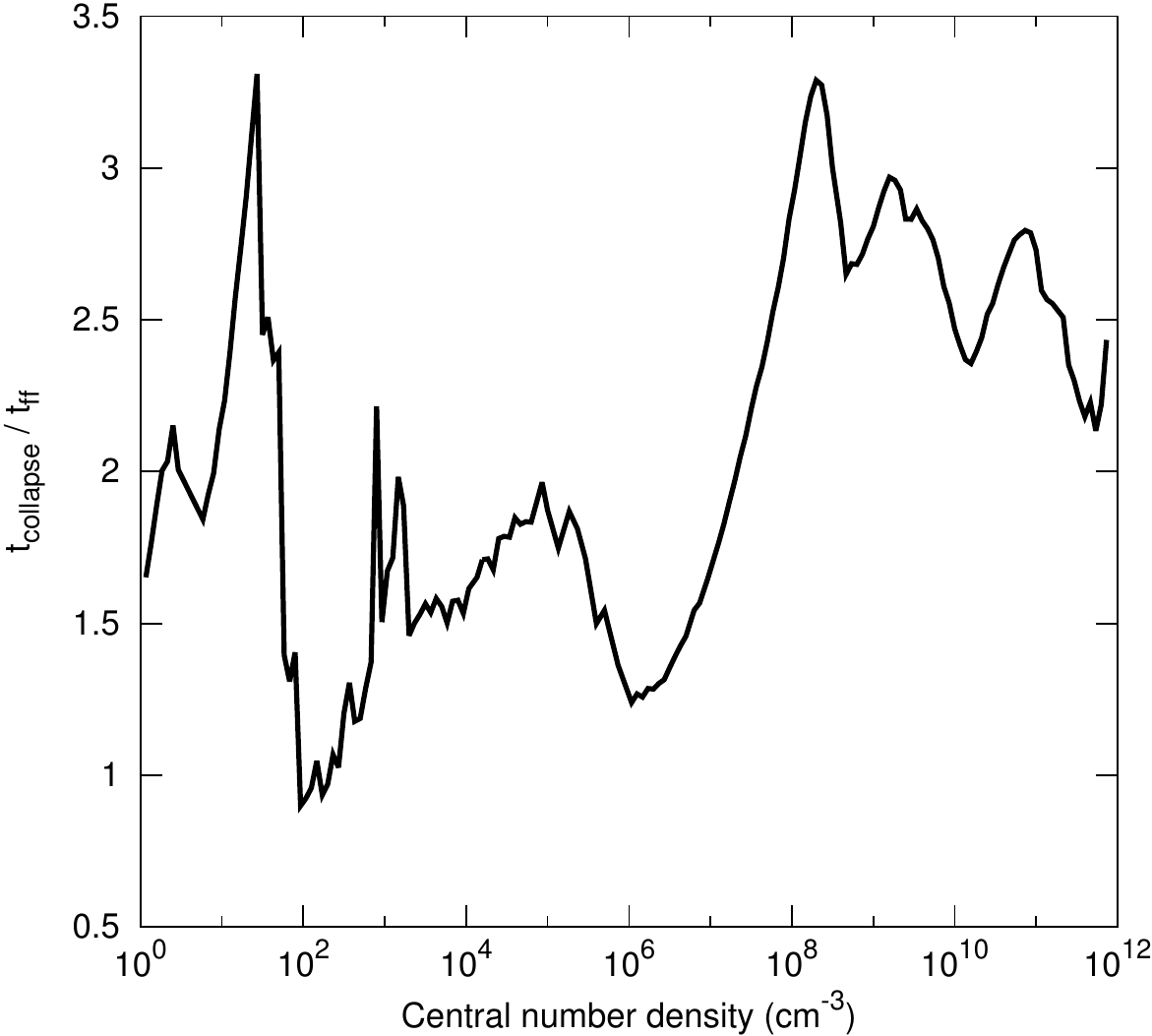}\\
\end{center}
{
{\bf Fig.~S7. Collapse time of the cloud center normalized by the free-fall time for Run-B.}
Around the virial radius (gas number density $n_{\rm H} \sim 10$\,cm$^{-3}$),
the gas collapse was initially prevented by the dynamical support owing to
the streaming motions.
After H$_2$-cooling became efficient, the cloud contracted quickly ($n_{\rm H} \sim 10^2$\,cm$^{-3}$).
The enclosed mass exceeded the effective Jeans mass when $n_{\rm H} \sim 10^5$\,cm$^{-3}$ (Fig.~S3C). 
Finally at $n_{\rm H} > 10^8$\,cm$^{-3}$,
three-body H$_2$ formation reactions and associated H$_2$-line cooling 
accelerated the cloud collapse. 
}
\end{figure}

\clearpage
\begin{figure}[!ht]
\begin{center}
\includegraphics[width=0.7\textwidth]{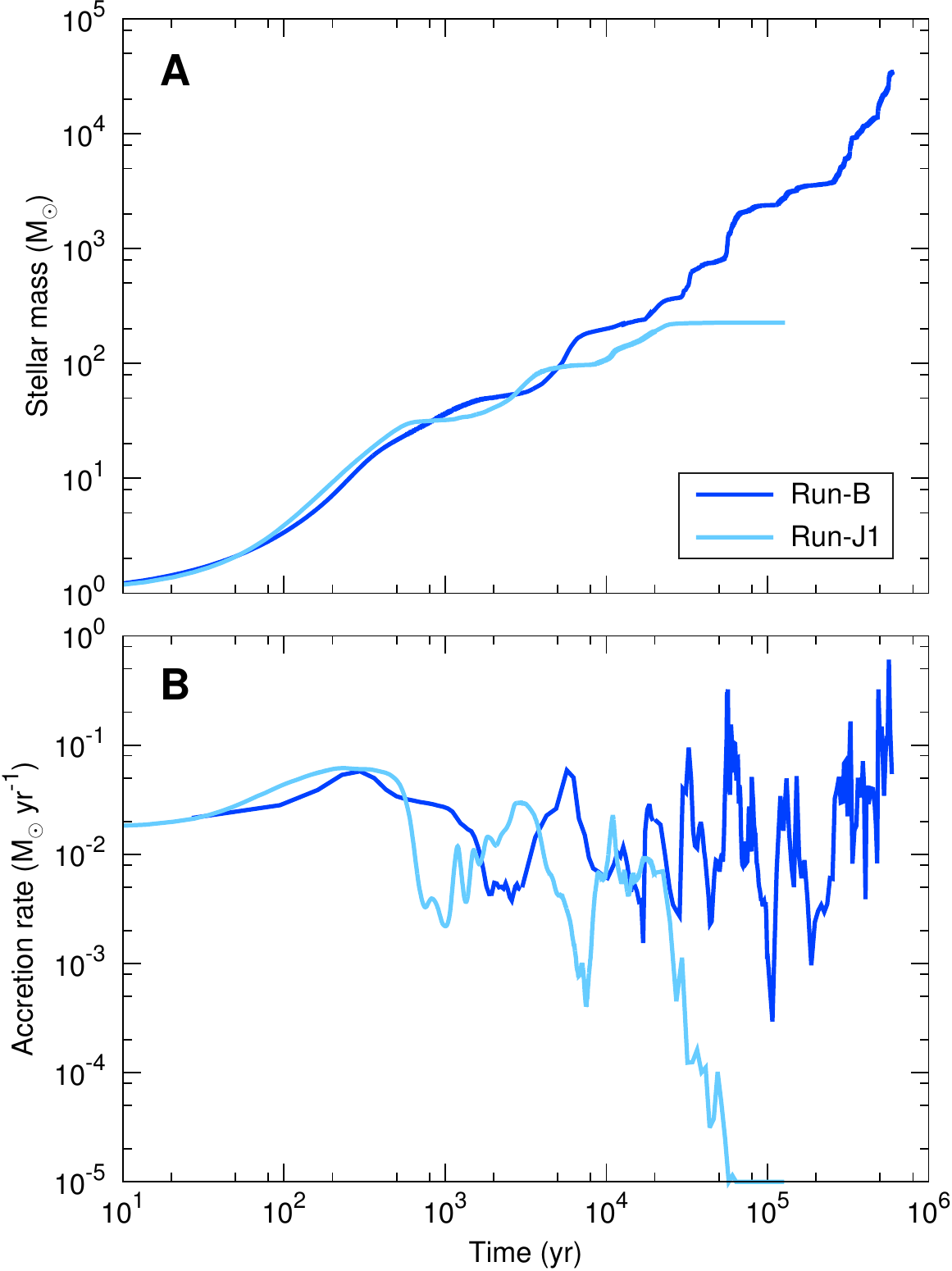}\\
\end{center}
{
{\bf Fig.~S8. Time evolution of stellar mass and accretion rate in Run-B and Run-J1 during the protostellar accretion phase.}
In Run-J1, the strong UV radiation forms an H~\textsc{ii} region and photo-evaporates the accreting gas when the stellar mass is $227\,M_\odot$.
}
\end{figure}

\clearpage
\begin{figure}[!ht]
\begin{center}
\includegraphics[width=0.7\textwidth]{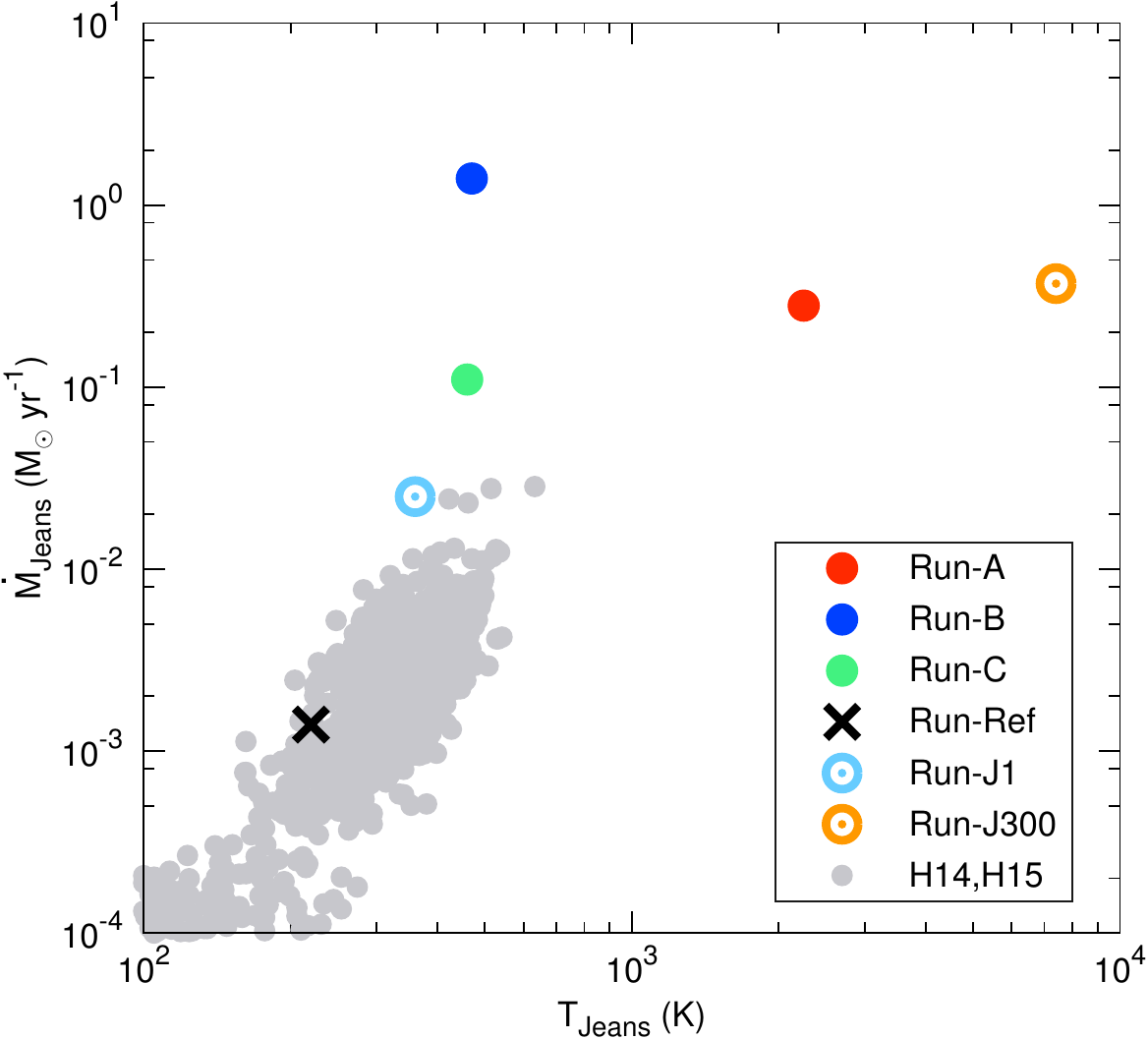}\\
\end{center}
{
{\bf Fig.~S9. Instantaneous mass infall rate within Jeans-unstable cloud $\dot{M}_{\rm Jeans}$ as a function of the gas temperature $T_{\rm Jeans}$.}
The filled circles show the results of runs with streaming velocities: Run-A (red), B (blue), and C (green).
The cross shows the reference simulation: Run-Ref, and
the open circles are for Run-J1 (cyan) and J300 (orange).
The dots show the simulation results of ordinary primordial star-formation simulations of H14\cite{hirano14} and H15\cite{hirano15a}.
For Run-A, although the temperature is lower than Run-J300, the mass infall rate is similar.
For Run-B and Run-C, the temperature is almost similar with Run-J1 but the infall rates are larger.
}
\end{figure}

\clearpage
\begin{table}[ht]
{
{\bf Table~S1. Summary of simulation results.}
The first column represents names of simulations.
The next two columns show the model parameters of the cosmological initial conditions:
the streaming velocity normalized by the root-mean-square value and
the normalized density fluctuation amplitude.
The next five columns summarize the properties of the primordial star-forming gas cloud in each simulation:
$z$ refers the collapse redshift,
$R_{\rm virial}$, $M_{\rm virial}$, and $V_{\rm virial}$ refer to
the radius, mass, and circular velocity of the host dark matter halo, and
$M_{\rm Jeans}$ refers to the Jeans mass when the gas cloud becomes gravitationally unstable (Fig.~S2).
The last two columns show the final state of the accreting protostellar calculations:
accretion time and final stellar mass.
}
\begin{center}
\begin{tabular}{lccrrrrrrrr}
\hline
Run & $v_{\rm bc}$ & $\sigma_8$ & $z$ & $R_{\rm virial}$ & $M_{\rm virial}$ & $V_{\rm virial}$ & $M_{\rm Jeans}$ & $t_{\rm acc}$ & $M_{\rm star}$ \\
& ($\sigma_{\rm bc}$) & $ $ & $ $ & (pc) & ($M_\odot$) & (km\,s$^{-1}$) & ($M_\odot$) & ($10^6$\,yr) & ($M_\odot$) \\
\hline\hline
A & $3$ & $2.0$ & $49.4$ & $93$ & $2.6 \times 10^7$ & $21.4$ & $146$,$000$ & $0.31$ & $100$,$000$ \\
B & $3$ & $1.2$ & $30.5$ & $171$ & $2.2 \times 10^7$ & $13.3$ & $26$,$000$ & $0.60$ & $34$,$000$ \\
C & $3$ & $1.2$ & $37.0$ & $101$ & $7.0 \times 10^6$ & $9.2$ & $2$,$000$ & $0.57$ & $4$,$400$ \\
Ref & $0$ & $0.8$ & $34.6$ & $26$ & $1.6 \times 10^5$ & $3.9$ & $400$ & & \\
J1 & $0$ & $1.2$ & $43.8$ & $46$ & $1.5 \times 10^6$ & $6.3$ & $1$,$300$ & $0.12$ & $227$ \\
J300 & $0$ & $1.2$ & $37.6$ & $86$ & $5.5 \times 10^6$ & $6.5$ & $20$,$000$ & & \\
\hline
\end{tabular}\\
\end{center}
\end{table}

\end{document}